\documentclass[preprint,aps,prc,tightenlines,nofootinbib,
showpacs,showkeys,superscriptaddress]{revtex4}
\begin{document}
\input{psfig}
\input{epsf}
\def\Im{\mbox{\sl Im\ }}
\def\pd{\partial}
\def\oln{\overline}
\def\olft{\overleftarrow}
\def\ds{\displaystyle}
\def\bgreek#1{\mbox{\boldmath $#1$ \unboldmath}}
\def\sla#1{\slash \hspace{-2.5mm} #1}
\newcommand{\bra}{\langle}
\newcommand{\ket}{\rangle}
\newcommand{\vep}{\varepsilon}
\newcommand{\met}{{\mbox{\scriptsize met}}}
\newcommand{\lab}{{\mbox{\scriptsize lab}}}
\newcommand{\cm}{{\mbox{\scriptsize cm}}}
\newcommand{\mcal}{\mathcal}
\newcommand{\Del}{$\Delta$}
\newcommand{\g}{{\rm g}}
\long\def\Omit#1{}
\long\def\omit#1{\small #1}
\def\beq{\begin{equation}}
\def\eeq{\end{equation} }
\def\bea{\begin{eqnarray}}
\def\eea{\end{eqnarray}}
\def\eqref#1{Eq.~(\ref{eq:#1})}
\def\eqlab#1{\label{eq:#1}}
\def\figref#1{Fig.~\ref{fig:#1}}
\def\figlab#1{\label{fig:#1}}
\def\tabref#1{Table \ref{tab:#1}}
\def\tablab#1{\label{tab:#1}}
\def\secref#1{Section~\ref{sec:#1}}
\def\seclab#1{\label{sec:#1}}
\def\VYP#1#2#3{{#1} (#2) #3}  
\def\NP#1#2#3{Nucl.~Phys.~\VYP{#1}{#2}{#3}}
\def\NPA#1#2#3{Nucl.~Phys.~A~\VYP{#1}{#2}{#3}}
\def\NPB#1#2#3{Nucl.~Phys.~B~\VYP{#1}{#2}{#3}}
\def\PL#1#2#3{Phys.~Lett.~\VYP{#1}{#2}{#3}}
\def\PLB#1#2#3{Phys.~Lett.~B~\VYP{#1}{#2}{#3}}
\def\PR#1#2#3{Phys.~Rev.~\VYP{#1}{#2}{#3}}
\def\PRC#1#2#3{Phys.~Rev.~C~\VYP{#1}{#2}{#3}}
\def\PRD#1#2#3{Phys.~Rev.~D~\VYP{#1}{#2}{#3}}
\def\PRL#1#2#3{Phys.~Rev.~Lett.~\VYP{#1}{#2}{#3}}
\def\FBS#1#2#3{Few-Body~Sys.~\VYP{#1}{#2}{#3}}
\def\AP#1#2#3{Ann.~of Phys.~\VYP{#1}{#2}{#3}}
\def\ZP#1#2#3{Z.\ Phys.\  \VYP{#1}{#2}{#3}}
\def\ZPA#1#2#3{Z.\ Phys.\ A\VYP{#1}{#2}{#3}}
\def\half{\mbox{\small{$\frac{1}{2}$}}}
\def\quarter{\mbox{\small{$\frac{1}{4}$}}}
\def\nn{\nonumber}
\newlength{\PicSize}
\newlength{\FormulaWidth}
\newlength{\DiagramWidth}
\newcommand{\vslash}[1]{#1 \hspace{-0.42 em} /}
\def\olaf{\marginpar{Mod-Olaf}}
\def\her{\marginpar{$\Longleftarrow$}}
\def\bel{\marginpar{$\Downarrow$}}
\def\abo{\marginpar{$\Uparrow$}}



\title{The Adler-Weisberger and Goldberger-Miyazawa-Oehme sum rules as probes
of constraints from analyticity and chiral symmetry in dynamical models
for pion-nucleon scattering}

\author{S. Kondratyuk}
\thanks{Corresponding author}
\email{kondra@physics.sc.edu}
\affiliation{Nuclear Theory Group, 
Department of Physics and Astronomy,
University of South Carolina,
712 Columbia, SC 29208, USA}
\author{K. Kubodera}
\affiliation{Nuclear Theory Group, 
Department of Physics and Astronomy,
University of South Carolina,
712 Columbia, SC 29208, USA}
\author{F. Myhrer}
\affiliation{Nuclear Theory Group, 
Department of Physics and Astronomy,
University of South Carolina,
712 Columbia, SC 29208, USA}
\author{O. Scholten}
\affiliation{Kernfysisch Versneller Instituut,
University of Groningen,
9747 AA Groningen, The Netherlands}


\begin{abstract}

The Adler-Weisberger and Goldberger-Miyazawa-Oehme sum rules are calculated
within a relativistic, unitary and crossing symmetric
dynamical model for pion-nucleon scattering using two different methods:
1) by evaluating the scattering amplitude at the corresponding
low-energy kinematics and 2) by evaluating the sum-rule integrals with
the calculated total cross section. 
The discrepancy between the results of the two
methods provides a measure of the breaking of analyticity and chiral symmetry 
in the model. The contribution of the $\Delta$ resonance, including
its dressing with meson loops, is discussed in some detail and found to be
small.

\end{abstract}




\maketitle


\section{Introduction} \seclab{intro}

It is known that much of the hadronic dynamics at low energies is 
determined by nucleon and meson exchanges while the intermediate-energy region is dominated by
various resonances. Chiral symmetry is an
important physical constraint that governs the
meson-nucleon interactions at low energies and unitarity is an essential
property at intermediate energies. 
A natural link between the low- and intermediate-energy regions is
provided by various sum rules.
In this paper we present a dynamical study of two sum rules for pion-nucleon scattering:
the Adler-Weisberger (AW) and Goldberger-Miyazawa-Oehme (GMO) 
sum rules~\cite{Adl65,Wei96,Gol55}. These relate 
integrals over the total cross section to the scattering amplitude calculated 
at a subthreshold kinematic point (in the case of the AW sum rule) and at threshold 
(in the case of the GMO sum rule), in the isospin-odd channel.

The sum rules follow from 
the properties of relativistic invariance, unitarity, analyticity and 
crossing symmetry (from whose combination, and assuming no subtractions, one derives a
dispersion relation for the isospin-odd amplitude), augmented by
chiral symmetry constraints.
The latter may be taken into account either  
by using the current algebra commutator for axial charges and
the PCAC relation for the divergence 
of the axial current~\cite{Adl65,Gol55} 
or by using a chirally invariant Lagrangian at tree level~\cite{Wei96}. 
In deriving the sum rules one assumes that due to the smallness of the pion mass $\mu$, 
the non-pole part of the amplitude is a slowly 
varying function of external pion four-momenta throughout the low-energy region of
other variables, thus retaining only the terms of the lowest order in $\mu$.
The higher orders corrections come from effects of finite nucleon size (by estimating
the pion-nucleon form factor~\cite{Bro71} or by evaluating pion loops~\cite{Bec01}) and from
resonance contributions (of which the $\Delta$ is expected to be dominant~\cite{Bro71}).

Being based on the very general physical principles,
the sum rules can serve as stringent
tests of dynamical models of pion-nucleon scattering. 
Suppose one obtains the low-energy quantities predicted by
the sum rules in two ways: first, from the calculated low-energy scattering amplitude
itself and, second, by integrating the calculated total cross sections with the
appropriate weight. In doing so, it is essential that the two
ways of evaluation (called here the ``low-energy" and ``sum-rule" evaluations) be done within
the {\em same} dynamical model. Then the results of the low-energy and sum-rule
evaluations would coincide provided i) the basic physical properties from which the sum 
rules are derived hold true, and ii) the model fulfils these properties exactly.
In any practical situation there will be a certain discrepancy between
the low-energy and the sum-rule results. This discrepancy allows one to quantify
violations of either one or both conditions i) and ii). Such a
study was first done for nucleon Compton scattering, where the
anomalous magnetic moment and
polarisabilities of the nucleon are related to the Gerasimov-Drell-Hearn and Baldin-Lapidus 
sum rules~\cite{Kon02p}. 

The purpose of the present paper is to
carry out a similar investigation for the Adler-Weisberger and Goldberger-Miyazawa-Oehme
sum rules. We use the ``Dressed K-matrix Model"~\cite{Kon02,Kon01,Kon00} which is unique in that
it provides a good description of pion-nucleon scattering in {\em both} 
intermediate- and low-energy regions, while also 
yielding the nucleon sigma-term in agreement with 
recent experimental analyses. The model obeys relativistic invariance, unitarity and 
crossing symmetry exactly, and incorporates constraints from analyticity and chiral symmetry 
approximately. We interpret the violation of analyticity and chiral symmetry 
in terms of meson-loop corrections to the
free nucleon and $\Delta$ propagators and to bare $\pi N N$ and $\pi N\Delta$ vertices.
The calculation of the loops is done up to infinite order within a consistent dressing procedure.
As a result of the present study,
we find that the dressing improves the agreement between the low-energy and sum-rule
calculations. We will focus on the effects of
the $\Delta$-resonance dressing in some detail.

\section{The Adler-Weisberger sum rule} \seclab{aw_sr}

The sum rules discussed in this paper are formulated in terms of the
isospin-odd amplitude $D^-(\nu)$ for forward 
pion-nucleon scattering.\footnote{Throughout the paper, 
the kinematical conventions and definitions for pion-nucleon scattering amplitude
follow those of Ref.~\cite{Hoh83}.} We denote the nucleon and pion masses as 
$m$ and $\mu$, respectively ($m=938$ MeV, $\mu=138$ MeV), and the
pion-nucleon coupling constant $g_{\pi N}=13.02$~\cite{Sto93,Kor98} 
(this value of $g_{\pi N}$ is also compatible with Refs.~\cite{Arn94}). 
In the following we shall use the invariant energy variable 
$\nu=(s-u)/(4m)$, with $s, u$ and $t$ being the Mandelstam variables, constrained by 
$s+u+t=2 m^2 + q^2 + q^{\prime 2}$, where $q$ and $q'$ denote the initial and final pion
four-momenta, respectively ($q^2=q^{\prime 2} = \mu^2$ for on-shell pions).
We shall also use 
$\omega = (s-m^2-q^2)/(2m)$, the energy of the incoming pion in the laboratory frame.
The total cross sections for the $\pi^- p$ and $\pi^+ p$
scattering processes
will be denoted as $\sigma_{\pi^- p}$ and $\sigma_{\pi^+ p}$, respectively.

The Adler-Weisberger (AW) sum rule for pion-nucleon scattering~\cite{Adl65,Wei96,Bro71} 
can be written as a relation between the integral over the weighted total
cross section and the low-energy limit of the isospin-odd scattering amplitude:
\beq
C_{AW} = I_{AW},
\eqlab{sr_aw}
\eeq
where the left-hand side represents a low-energy limit of
the forward scattering amplitude: 
\beq
C_{AW} = 2 F_\pi^2 \,
\lim_{\nu \rightarrow 0} 
\left\{ \frac{{D}^-(\nu)}{\nu} + \frac{g^2_{\pi N}}{2 m^2} \right\},
\eqlab{c_aw}
\eeq
and the right-hand side is given by the sum-rule integral 
\beq
I_{AW} = 2 F_\pi^2 \left\{ \frac{1}{\pi}\int_{\mu}^{\infty}\!d \omega \, 
\frac{\sqrt{\omega^2-\mu^2}}{\omega^2} \, 
\left[ \sigma_{\pi^- p}(\omega) - \sigma_{\pi^+ p}(\omega) \right] +
\frac{g^2_{\pi N}}{2 m^2} \right\},
\eqlab{int_aw}
\eeq
where $F_\pi=92.4$ MeV is the pion decay constant.
The presence of the factor $2 F_\pi^2$ and of the nucleon pole contribution 
$g_{\pi N}^2/(2 m^2)$ in 
Eqs.~(\ref{eq:c_aw}) and (\ref{eq:int_aw}) is a matter of convenience since 
for a vanishing pion mass chiral symmetry dictates that~\cite{Adl65,Wei96}
\beq
C_{AW} = 1 + {\mathcal{O}}(\mu^2/m^2).
\eqlab{caw_let}
\eeq
Thus the dominant correction to \eqref{sr_aw} is of the order $\mathcal{O}(\mu^2/m^2)$, reflecting
the fact that the physical pion has a non-vanishing mass whereas the AW 
sum rule is derived formally for the scattering of a massless pion.

\section{The Goldberger-Miyazawa-Oehme sum rule} \seclab{gmo_sr}

The Goldberger-Miyazawa-Oehme (GMO) sum rule~\cite{Gol55,Bro71,Eri02} can be written in the form
\beq
C_{GMO} = I_{GMO},
\eqlab{sr_gmo}
\eeq
where the low-energy part is determined by the isospin-odd scattering length $a^-$:
\beq
C_{GMO} = 4 \pi \mu a^- ,
\eqlab{c_gmo}
\eeq
and the sum-rule integral is given by
\beq
I_{GMO} = \mu^2 \left\{
\frac{1}{\pi}\int_{\mu}^{\infty}\!d \omega \,  
\frac{1}{\sqrt{\omega^2-\mu^2}}
\left[ \sigma_{\pi^- p}(\omega) - \sigma_{\pi^+ p}(\omega) \right] +
\frac{g^2_{\pi N}}{2 m^2} \right\}. 
\eqlab{int_gmo}
\eeq
Similar to $C_{AW}$ in \eqref{c_aw}, $C_{GMO}$ is related to a low-energy value of the
scattering amplitude since $a^-$ is proportional to $D^-$ at threshold:
\beq
a^- = \frac{D^-(\nu=\mu)}{4 \pi (1+\mu/m)}. 
\eqlab{length}
\eeq
The last equation also shows that the GMO sum rule \eqref{sr_gmo} is written up to 
higher-order terms of $\mathcal{O}(\mu/m)$. This could be compared with the accuracy 
$\mathcal{O}(\mu^2/m^2)$ of the AW sum rule \eqref{sr_aw}, 
as indicated in \eqref{caw_let}.

It should be pointed out that some terms in the sum rules might formally appear of 
a lower order in $\mu/m$ than they actually are: for example, the leading
terms in \eqref{sr_gmo} might be interpreted as being of $\sim{{\mathcal{O}}(\mu^2/m^2})$, but the
explicit evaluation shows that, due to large accompanying coefficients, 
they are comparable with unity (see \secref{results}). 
By the same token, a proper care should be taken in estimating the leading corrections to
the sum rules Eqs.~(\ref{eq:sr_aw}) and (\ref{eq:sr_gmo}). A comprehensive analysis of the
intrinsic accuracy of the sum rules was presented recently in Ref.~\cite{Eri02}, where
the GMO sum rule was utilised to deduce the pion-nucleon coupling constant.
For the purposes of the present paper, the precise value of $g_{\pi N}$ is not very important
since we focus on the comparison of the low-energy and sum-rule evaluations;
it is essential however that we keep all parameters (including $g_{\pi N}$) unchanged
in evaluating both left- and right-hand sides of Eqs.~(\ref{eq:sr_aw}) and (\ref{eq:sr_gmo}).

\section{Basic features of the model}\seclab{model}

The objective of our calculation is to check the validity of the AW and GMO sum rules
using a dynamical model for pion-nucleon interaction applicable at low and 
intermediate energies. The crucial point is that the evaluation of the
left- and right-hand sides of Eqs.~(\ref{eq:sr_aw}) and (\ref{eq:sr_gmo}) is done here 
in the {\em same} model, rather than utilising one approach to evaluate the
low-energy limits of the amplitudes (the left-hand sides) and another approach to evaluate the
integrals (the right-hand sides).
Only if Eqs.~(\ref{eq:sr_aw}) and (\ref{eq:sr_gmo}) are calculated entirely within the 
same model will their validity reflect the extent to which the 
basic symmetries are fulfilled in the chosen model, rather than stem from 
(possible) incompatibility of different approaches. 

We use the ``Dressed K-matrix Model" whose detailed description can be found in
Refs.~\cite{Kon02,Kon01,Kon00}. Here we recapitulate only
the principal features of the approach. The properties of relativistic invariance, 
(two-body) unitarity and crossing symmetry are fulfilled in the model exactly, while
those of chiral symmetry and analyticity are implemented partially. 

Chiral symmetry constraints
are effectively incorporated at low energies through the use of a predominantly pseudovector
pion-nucleon coupling in the s- and u-channel nucleon exchange diagrams, plus
$\rho$ and $\sigma$ t-channel exchanges with parameters fitted to provide a good
low-energy description of the s- and p-wave phase shifts~\cite{Kon01,Kon00}. 
Effects of the explicit violation
of chiral symmetry appear due to the finiteness of the pion mass. In the context of the AW 
sum rule, these effects manifest themselves in the deviation of $C_{AW}$ from unity, as shown in
\eqref{caw_let}. 

Analyticity is implemented in the model through the dressing procedure for the $\pi N N$ and
$\pi N \Delta$ vertices and the nucleon and $\Delta$ propagators. These three- and two-point
functions are calculated as a solution of a system of coupled integral equations, which 
amounts to including meson loop corrections up to infinite order
(for a specific description of the class of the loops generated by the dressing, 
see Refs.~\cite{Kon02,Kon00}).
The solution is based on
the use of cutting rules and dispersion relations in an iteration procedure, which 
allows us to obtain analytic two- and three-point functions (i.e.~the dressed propagators and
vertices, respectively). The scattering amplitude is obtained in a
K-matrix framework, where the K matrix is constructed from skeleton diagrams built out of 
the dressed vertices and propagators. As a result of the dressing, 
analyticity is partially incorporated into the K-matrix 
framework (as opposed to the traditional K-matrix models where analyticity is strongly violated
because of K matrices built out of tree diagrams).
The full restoration of analyticity could be achieved by
dressing the four-point irreducible functions in the amplitude on the same footing with the
presently dressed two- and three-point functions (see discussions in
Refs.~\cite{Kon02p,Kon02,Kon01,Kon00}).
 
The calculation presented in this paper provides a useful means of quantifying the
extent to which analyticity and chiral symmetry are broken in the model. Indeed, since
the sum rules are based on the combination of relativistic invariance, unitarity, crossing symmetry,
analyticity and chiral symmetry, and only the latter two properties are not fully implemented 
in the model, the difference between the evaluation of the low-energy amplitudes (represented by
the left-hand sides of of Eqs.~(\ref{eq:sr_aw}) and (\ref{eq:sr_gmo})) and the evaluation of
the sum-rule integrals (the right-hand sides of Eqs.~(\ref{eq:sr_aw}) and (\ref{eq:sr_gmo})) will
serve as a measure of the violation of analyticity and chiral symmetry constraints. 
Such a comparison of low-energy and sum-rule evaluations was carried out earlier for
the Baldin-Lapidus, Gerasimov-Drell-Hearn and other related sum rules in Compton 
scattering~\cite{Kon02p}. The main difference between the present case of the AW and GMO sum rules
and that of the sum rules in Compton scattering is that now we test 
not only the violation of analyticity
but also effects of the explicit chiral symmetry breakdown.

\section{Results of the calculation} \seclab{results}

We will study the effects of the dressing of the two- and three-point functions 
by comparing three different calculations, called the ``Dressed", ``Bare" and
``Dressed, but bare $\Delta$" calculations. The ``Bare" calculation corresponds to
the use of free propagators and bare vertices in the K matrix, whereas the ``Dressed" one 
to the use of the fully dressed vertices and propagators.
The ``Bare" calculation is equivalent to a traditional K-matrix model with a tree-level
K matrix; therefore, the violation of analyticity is maximal in this case. 
Analyticity is partially restored 
in the ``Dressed" calculation since the dressing of the two- and 
three-point functions incorporates the use of dispersion relations. 
The remaining violation of analyticity comes mainly from the lack of 
the dressing of the four-point contact diagrams.
The ``Dressed, but bare $\Delta$" calculation
highlights effects of the dressing of the $\Delta$ resonance and will be described in the
next subsection.
Since we want to consider the genuine effects of the dressing, 
the same set of parameters are used in the ``Dressed", ``Bare" and ``Dressed, but bare $\Delta$" 
calculations (all coupling constants, including the masses,
the coupling constants and the cut-offs of the bare form factors, are given in Ref.~\cite{Kon02}).

To investigate convergence of the sum rules, we calculate the integrals
in Eqs.~(\ref{eq:int_aw}) and (\ref{eq:int_gmo}) up to a pion energy $\omega_{\mbox{up}}$.
The dependence of the calculated AW and GMO sum rules 
on $\omega_{\mbox{up}}$ is shown in 
Figs.~\ref{fig:sr_aw} and \ref{fig:sr_gmo}, respectively.
\begin{figure}
\centerline{{\epsfxsize 10cm \epsffile[40 70 600 570]{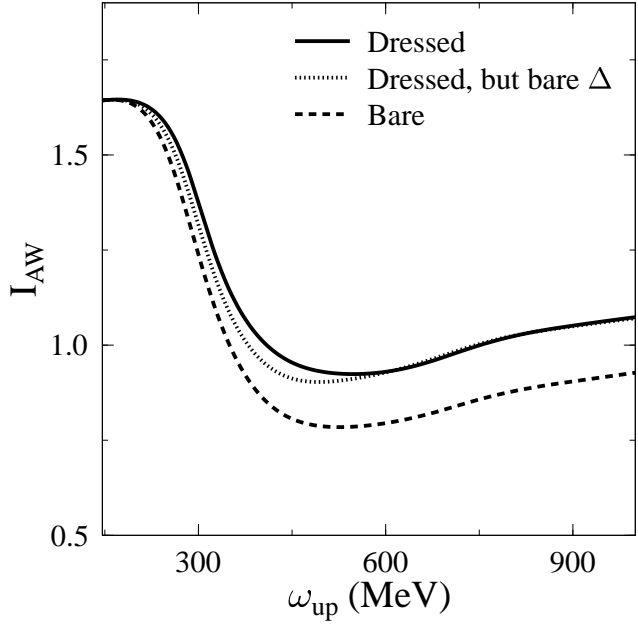}}}
\caption[f1]{
Dependence of the AW sum-rule integral \eqref{int_aw} on the upper limit of integration.
\figlab{sr_aw}} 
\end{figure}
\begin{figure}
\centerline{{\epsfxsize 10cm \epsffile[40 70 600 520]{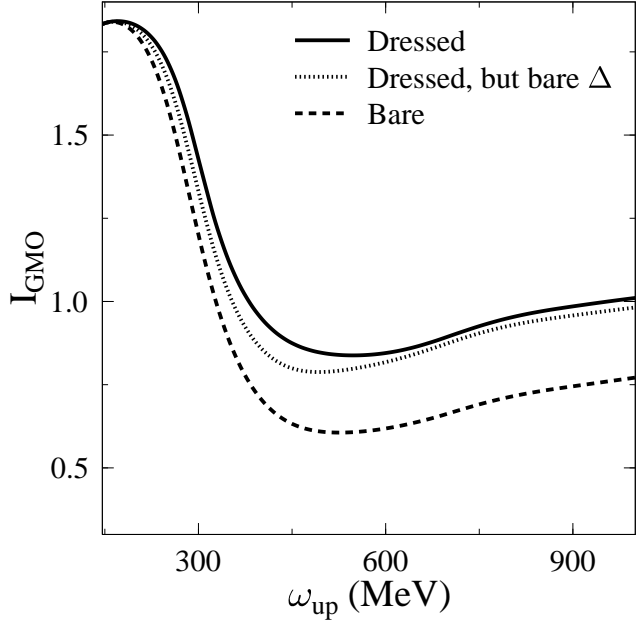}}}
\caption[f2]{
The GMO sum-rule integral \eqref{int_gmo} as a function of the upper limit of integration.
\figlab{sr_gmo}}
\end{figure}
Even though a full convergence is not achieved by $\omega_{\mbox{up}} \approx 1000$ MeV, 
the major features of the sum-rules are obtained due to the nucleon and the $\Delta$-resonance,
as can be seen from \figref{xs_tot} where we show the calculated total cross sections.
\begin{figure}
\centerline{{\epsfxsize 10cm \epsffile[30 70 600 690]{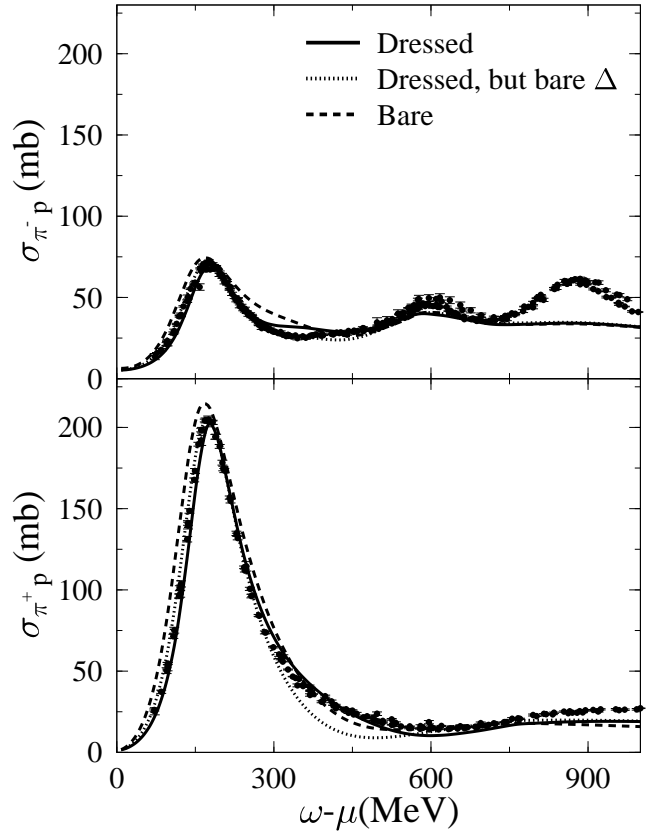}}}
\caption[f3]{
Total cross sections for the $\pi^+ p$ and $\pi^- p$ scattering processes
calculated in the model.
The experimental points are taken from the CNS Data Analysis 
Centre~\cite{CNS01}. 
\figlab{xs_tot}}
\end{figure}
The hump in the cross sections around $\omega - \mu \approx 900$ MeV 
is not reproduced since we did not include the D15 and F15 resonances
in the model at present. We checked however that this higher-energy feature 
of the cross section is suppressed by the weight factors in 
Eqs.~(\ref{eq:int_aw}) and (\ref{eq:int_gmo}). 
This observation is in complete agreement with 
the conclusion of Ref.~\cite{Bro71} that the resonances heavier than the $\Delta$ have a negligible
effect on the sum rules. 

We extract the values of the sum rules at two representative 
pion energies, $\omega_{\mbox{up}}=750$ MeV and $\omega_{\mbox{up}}=1000$ MeV, and regard 
them as the results of the sum-rule evaluation,
i.e.~as $I_{AW}$ and $I_{GMO}$ given by Eqs.~(\ref{eq:int_aw}) and (\ref{eq:int_gmo}), 
respectively.  The low-energy quantity $C_{GMO}$ was calculated at threshold, according to
\eqref{c_gmo}, and the low-energy values of $C_{AW}$, as given by
\eqref{c_aw}, were calculated at two subthreshold points: 
$\{ s=u=m^2, t=2 \mu^2 \}$ and $\{ s=u=m^2, t=0 \}$. 
The former is the Cheng-Dashen point and the latter
is closely related to the Weinberg point, except that at the Weinberg point both pions have 
a zero four-momentum squared whereas in our case the pions are on-shell.
All these values of $C$ and $I$ are given in \tabref{lesr_comp}.
\begin{table}
\caption[t1]{The values of the 
Adler-Weisberger and Goldberger-Miyazawa-Oehme sum rules evaluated
from the amplitude at the low-energy kinematical points and from the integrals over
total cross sections, 
as given by Eqs.~(\ref{eq:c_aw}), (\ref{eq:int_aw}),  
(\ref{eq:c_gmo}) and (\ref{eq:int_gmo}). The different low-energy values $C$ and 
sum-rule values $I$ are described in the text, as are the three
calculations with varying amounts of dressing.}
\begin{center}
\begin{tabular}{l c c c }
\hline\hline
 & \hspace*{1mm} Dressed  \hspace*{1mm} & 
   \hspace*{1mm} Dressed, but bare $\Delta$  \hspace*{1mm} & 
   \hspace*{1mm} Bare  \hspace*{1mm} \\
\hline
$C_{AW}(s=u=m^2,t=2\mu^2)$                    & $1.16$  & $1.14$ & $1.31$ \\ 
\hline
$C_{AW}(s=u=m^2,t=0)$                         & $1.10$  & $1.10$ & $1.25$ \\
\hline
$I_{AW}(\omega_{\mbox{up}}=750 \,\mbox{MeV})$         & $1.00$  & $1.00$ & $0.85$ \\
\hline 
$I_{AW}(\omega_{\mbox{up}}=1000 \,\mbox{MeV})$        & $1.08$  & $1.08$ & $0.93$ \\
\hline \hline
$C_{GMO}                 $                    & $1.10$  & $1.10$ & $1.22$ \\ 
\hline
$I_{GMO}(\omega_{\mbox{up}}=750 \,\mbox{MeV})$        & $0.92$  & $0.90$ & $0.67$ \\
\hline 
$I_{GMO}(\omega_{\mbox{up}}=1000 \,\mbox{MeV})$       & $1.01$  & $0.98$ & $0.77$ \\
\hline\hline
\end{tabular}
\end{center}
\tablab{lesr_comp}
\end{table}

The differences between
the the values of $C$ and $I$ are summarised in \tabref{lesr_diff}.
\begin{table}
\caption[t2]{Comparison of the low-energy and sum-rule evaluations of the 
quantities defined in Eqs.~(\ref{eq:c_aw}), (\ref{eq:int_aw}), (\ref{eq:c_gmo}),
(\ref{eq:int_gmo}) and given in \tabref{lesr_comp}. 
The differences are related to the violation of analyticity and 
the breaking of chiral symmetry.  
}
\begin{center}
\begin{tabular}{l c c c }
\hline\hline
 & \hspace*{1mm} Dressed  \hspace*{1mm} & 
   \hspace*{1mm} Dressed, but bare $\Delta$  \hspace*{1mm} & 
   \hspace*{1mm} Bare  \hspace*{1mm} \\
\hline
$C_{AW}(s=u=m^2,t=2\mu^2)-I_{AW}(\omega_{\mbox{up}}=750 \,\mbox{MeV})$ 
& $0.16$  & $0.14$ & $0.46$ \\ 
\hline
$C_{AW}(s=u=m^2,t=2\mu^2)-I_{AW}(\omega_{\mbox{up}}=1000 \,\mbox{MeV})$                         
& $0.08$  & $0.06$ & $0.38$ \\
\hline
$C_{AW}(s=u=m^2,t=0)-I_{AW}(\omega_{\mbox{up}}=750 \,\mbox{MeV})$         
& $0.10$  & $0.10$ & $0.40$ \\
\hline 
$C_{AW}(s=u=m^2,t=0)-I_{AW}(\omega_{\mbox{up}}=1000 \,\mbox{MeV})$        
& $0.02$  & $0.02$ & $0.32$ \\
\hline \hline
$C_{GMO}-I_{GMO}(\omega_{\mbox{up}}=750 \,\mbox{MeV})$                
& $0.18$  & $0.20$ & $0.55$ \\ 
\hline
$C_{GMO}-I_{GMO}(\omega_{\mbox{up}}=1000 \,\mbox{MeV})$        
& $0.09$  & $0.02$ & $0.45$ \\
\hline\hline
\end{tabular}
\end{center}
\tablab{lesr_diff}
\end{table}
The significance of these differences should be assessed in the context of the intrinsic
accuracy of the sum rules, which is $\sim \mathcal{O}(\mu^2/m^2) \sim \mathcal{O}(0.02)$ 
for the AW sum rule and $\sim \mathcal{O}(\mu/m) \sim \mathcal{O}(0.15)$ for the GMO sum rule,
as discussed in Sections \ref{sec:aw_sr} and \ref{sec:gmo_sr}. 
We see that the agreement comparable with the intrinsic
accuracy is achieved in the ``Dressed' calculation when comparing
the low-energy evaluation and the sum-rule evaluation 
with $\omega_{\mbox{up}}=1000$ MeV. By contrast, in the ``Bare" calculation 
the discrepancy between the low-energy and sum-rule evaluations is about three to ten times
larger than the intrinsic accuracy. 
This shows the importance of the dressing, mainly because it improves 
analyticity properties of the amplitudes while
maintaining unitarity and crossing symmetry.

\subsection{Effects of the dressing of the $\pi N \Delta$ vertex and $\Delta$ self-energy}
\seclab{deldr_effect}

Since leading corrections to the sum rules are expected to be due to the
$\Delta$ resonance~\cite{Bro71}, we consider the effects of the $\Delta$ dressing separately.
The calculation denoted ``Dressed, but bare $\Delta$" is based on a K matrix with
the free $\Delta$ propagator and the bare $\pi N \Delta$ vertex 
while all the other vertices and propagators 
remain dressed as in the full (i.e.~``Dressed") calculation.

The $\Delta$ dressing is described in detail in Ref.~\cite{Kon02}; here we will
mention only a few features and present the results.
The $\pi N \Delta$ vertex has the following structure:
\beq
V_\alpha(k,p) \sim G_{\Delta}(p^2)\,(k_\alpha \vslash{p} - p \cdot k \gamma_\alpha),
\eqlab{pnd_struc}
\eeq
where $p$ and $k$ denote the four-momenta of the $\Delta$ and pion, respectively,
and $G_\Delta(p^2)$ is the $\pi N \Delta$ form factor.
This vertex has the property $p \cdot V = 0$,
which ensures that only the physical spin-$3/2$ part is retained in 
the $\Delta$ propagator~\cite{Pas98}
\beq
S^{\alpha \beta}_{\Delta}(p) = \frac{\mathcal{P}_{3/2}^{\alpha
\beta}(p)}{\vslash{p}-m_{\Delta}-\Sigma_{\Delta}(p)},
\eqlab{del_prop}
\eeq
where $\mathcal{P}^{\alpha \beta}(p)$ is the projector on the spin-$3/2$ states,
$m_{\Delta}=1232$ MeV is the mass of the $\Delta$ resonance and
the $\Delta$ self-energy is given by 
\beq
\Sigma_\Delta(p) = A_\Delta(p^2) \vslash{p} + B_\Delta(p^2) m_{\Delta}-
(Z^\Delta_2-1)(\vslash{p}-m_{\Delta})-Z^\Delta_2\,\delta m_{\Delta} \,.
\eqlab{delself}
\eeq
The invariant self-energy functions $A_\Delta(p^2)$ and $B_\Delta(p^2)$ describe the
pion loop contribution, and the renormalisation constants $Z^\Delta_2$ and $\delta m_{\Delta}$
are fixed to ensure the correct pole location and residue of the dressed propagator.  
The $\pi N \Delta$ vertex gets dressed with an infinite number of
meson loops, i.e.~a starting bare form factor $G^0_\Delta(p^2)$ changes into $G_\Delta(p^2)$.
Correspondingly, the vertex \eqref{pnd_struc} changes its normalisation from a bare coupling 
constant to a physical one, the former being adjusted so that the dressed vertex yields 
the physical $\Delta \longrightarrow \pi N$ decay rate. 
It is important that the $\Delta$ self-energy $\Sigma_{\Delta}(p)$ and 
dressed $\pi N \Delta$ vertex are calculated 
simultaneously with the nucleon self-energy and dressed $\pi N N$ vertex;
all these two- and three-point functions are obtained as a solution of
one system of coupled integral equations (see Ref.~\cite{Kon02} for details).

The results of the ``Dressed, but bare $\Delta$" calculation, presented
in Figs.~\ref{fig:sr_aw}, \ref{fig:sr_gmo} and in 
Tables~\ref{tab:lesr_comp}, \ref{tab:lesr_diff},
show that the $\Delta$ dressing has a small effect on the AW and GMO sum rules.
Thus while the full (``Dressed") calculation is clearly superior to the ``Bare" one,
the $\Delta$ dressing considered in separation 
does not improve the agreement between the low-energy and sum-rule evaluations.
However, the dressed and bare $\pi N \Delta$ form factors 
differ noticeably, as shown in \figref{pnd_ff}.
At the $\Delta$ mass, $p^2 \approx 1.52$ GeV$^2$, 
the physical form factor is normalised to one by adjusting the normalisation of the bare 
form factor.
\begin{figure}
\centerline{{\epsfxsize 10cm \epsffile[25 125 540 620]{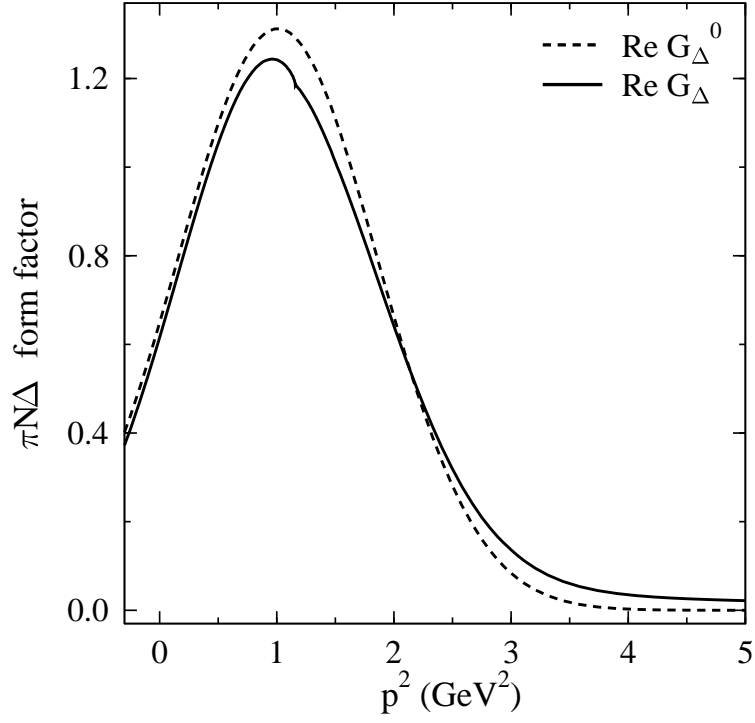}}}
\caption[f4]{
The bare (superscript $0$) and dressed $\pi N \Delta$ form factors, defined in \eqref{pnd_struc},
as functions of the four-momentum squared of the $\Delta$.
\figlab{pnd_ff}}
\end{figure}
The $\Delta$ self-energy
has a nontrivial structure as shown in \figref{del_se}. 
\begin{figure}
\centerline{{\epsfxsize 10cm \epsffile[25 40 540 490]{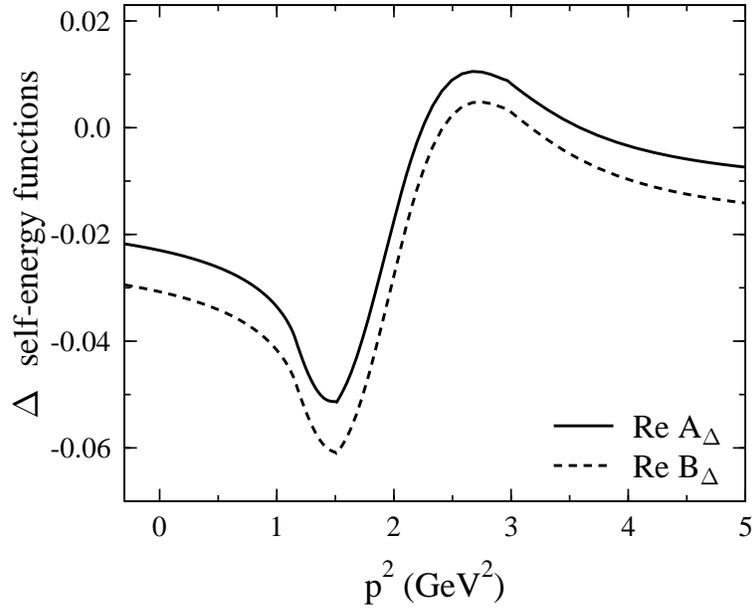}}}
\caption[f5]{
The $\Delta$ self-energy functions, defined in \eqref{delself}.
\figlab{del_se}}
\end{figure}
The effects of the $\Delta$ dressing manifest themselves 
quite prominently in the $P33$ pion-nucleon phase shift, 
shown in \figref{p33_deldr}, which is known to be
determined mainly by the $\Delta$.
\begin{figure}
\centerline{{\epsfxsize 16cm \epsffile[-20 210 580 560]{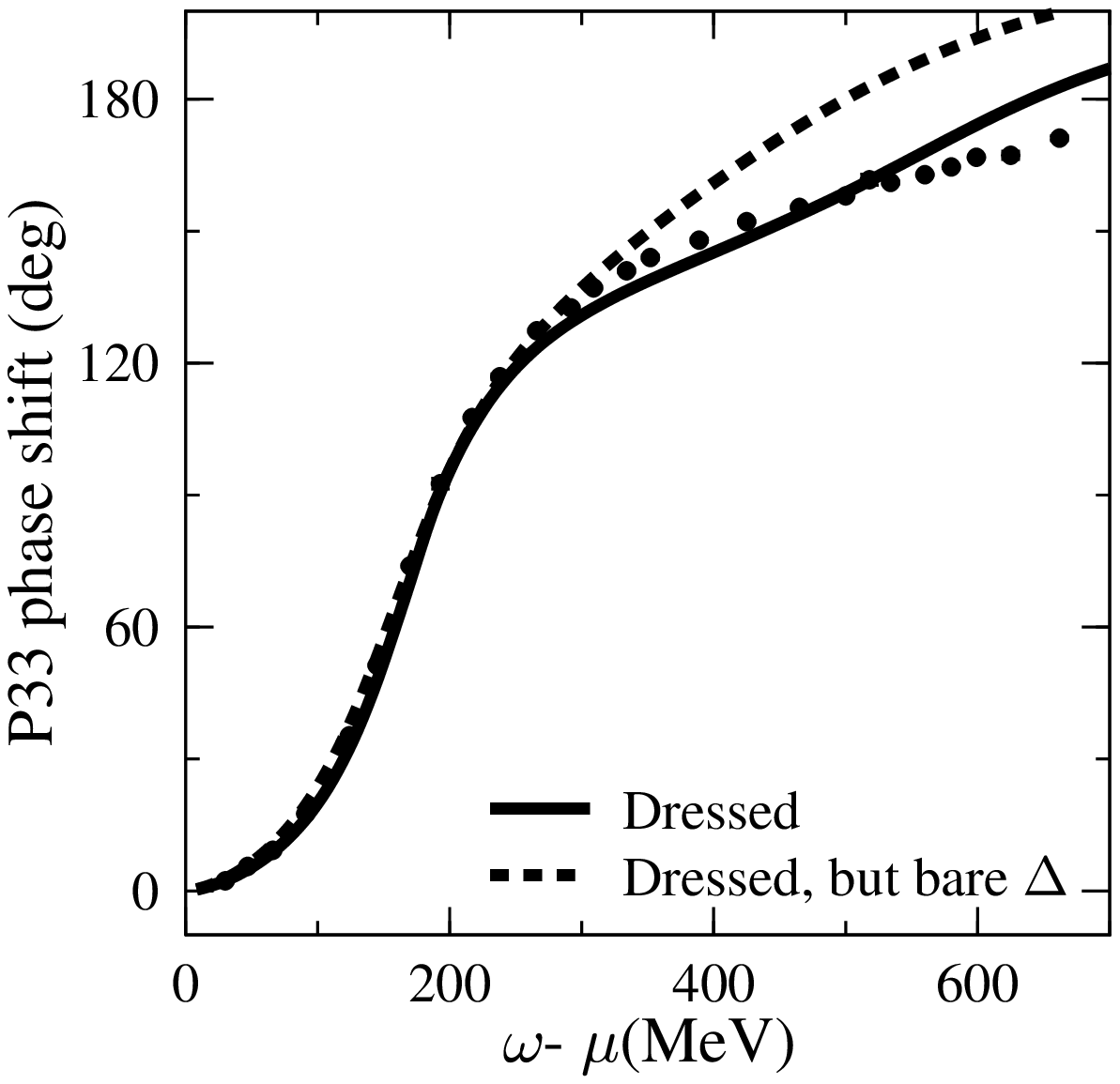}}}
\caption[f6]{
The effect of the $\Delta$ dressing on the P33 phase shift.  
The data points are from the CNS Data Analysis Centre~\cite{CNS01}.
\figlab{p33_deldr}}
\end{figure}
As mentioned above, we use the same set of parameters in the 
``Dressed", ``Bare" and ``Dressed, but bare $\Delta$" 
calculations. The only exception is made in presenting the $P33$ phase shift, where
we normalised the bare and physical $\pi N \Delta$ form factor to the same value at the $\Delta$ 
mass to ensure that
both curves in \figref{p33_deldr} pass through 90 degrees with the correct slope
at $\omega - \mu \approx 190$ MeV.

\section{Concluding remarks} \seclab{concl}

It is generally accepted that one can derive the 
Adler-Weisberger and Goldberger-Miyazawa-Oehme sum rules from 
relativistic invariance, unitarity, analyticity, crossing symmetry and chiral symmetry constraints.
Then the difference between the low-energy and the sum-rule calculations (i.~e.~between
the quantities $C$ and $I$) 
is a quantitative measure of the extent to which these properties are violated in 
the present model. 
Since relativistic invariance, two-body unitarity and crossing are exact
in the model, the discrepancy between the low-energy and sum-rule evaluations
is due to the partial fulfilment of analyticity and chiral symmetry. 

The dressing certainly improves analyticity properties 
of the amplitude, although as explained in \secref{model},
it does not restore analyticity completely. 
Chiral symmetry is explicitly broken in the model due to the
finiteness of the pion mass. This is reflected in the difference 
$C_{AW}(s=u=m^2,t=2\mu^2) - C_{AW}(s=u=m^2,t=0)$. Notably, the modulus of
this difference is of the same order as the discrepancy between either 
$C_{AW}(s=u=m^2,t=2\mu^2)$ or $C_{AW}(s=u=m^2,t=0)$ 
and the sum-rule values $I_{AW}$. It is also comparable with the measure of
convergence of the sum-rule integral, i.e.~with 
$I_{AW}(\omega_{\mbox{up}}=750 \,\mbox{MeV}) - I_{AW}(\omega_{\mbox{up}}=1000 \,\mbox{MeV})$.
Thus the extent to which the dressing affects the chiral symmetry constraints
is more difficult to quantify than that of analyticity. Nevertheless, our calculations indicate
that the explicit chiral symmetry breaking has a small influence on the 
AW and GMO sum rules. We have also shown that
the corrections due to the $\Delta$ resonance, including its dressing, 
are strongly suppressed.

It would be interesting to conduct comparisons between low-energy and sum-rule evaluations
within other approaches to pion-nucleon scattering, such as 
the Bethe-Salpeter equation~\cite{Lah99} or its reductions
(see, e.g.~\cite{Pas00}), traditional K-matrix models~\cite{Feu98,Kor98} or approaches based on
chiral Lagrangians (see, e.g.,~\cite{Lut02,Bec01} and references therein). Such a comparison
would however be meaningful only if, similarly to the
``Dressed K-matrix Model" used in the present study, 
a chosen approach is applicable at both low and
intermediate energies because one should be able to calculate reliably
both the low-energy amplitude and the total cross sections in the {\em same} framework.

\begin{acknowledgments}

S.~K., K.~K.~and F.~M.~are supported in part by the US National Science Foundation's
Grant No.~PHY-0140214, 
and the work of O.~S.~is part of the research program of the ``Stichting voor
Fundamenteel Onderzoek der Materie" (FOM) with financial support
from the ``Nederlandse Organisatie voor Wetenschappelijk
Onderzoek" (NWO).

\end{acknowledgments}


\end{document}